\documentclass[11pt]{article}
\usepackage{latexsym,epic,eepic}
\usepackage{amsmath,amsfonts}
\usepackage{epsfig}

\newtheorem{theorem}{Theorem}[section]
\newtheorem{corollary}[theorem]{Corollary}
\newtheorem{lemma}[theorem]{Lemma}

\newcommand{\qed}{\hfill$\diamond$}
\newcommand{\pf}{{\bf Proof: }}

\begin{document}

\title{Duality for Min-Max Orderings \\ and \\Dichotomy for Min Cost
Homomorphisms}

\date{}
\author{ Pavol Hell and Arash Rafiey \\
School of Computing Science \\ Simon Fraser University\\ Burnaby,
B.C., Canada, V5A 1S6
\thanks{ pavol@cs.sfu.ca , arashr@cs.sfu.ca}
} \maketitle

\begin{abstract}
Min-Max orderings correspond to conservative lattice polymorphisms. Digraphs with
Min-Max orderings have polynomial time solvable minimum cost homomorphism
problems. They can also be viewed as digraph analogues of proper interval graphs
and bigraphs.

We give a forbidden structure characterization of digraphs with a Min-Max ordering
which implies a polynomial time recognition algorithm. We also similarly characterize
digraphs with an extended Min-Max ordering, and we apply this characterization to
prove a conjectured form of dichotomy for minimum cost homomorphism problems.

\end{abstract}

\section{Introduction}

Let $H$ be any digraph. A linear ordering $<$ of $V(H)$ is a {\em Min-Max ordering}
if $i<j, s<r$ and $ir, js \in A(H)$ imply that $is \in A(H)$ and $jr \in A(H)$.

Min-Max orderings correspond to a particular type of lattice polymorphisms \cite{cater}.
For digraphs $G$ and $H$, a mapping $f:\ V(G)\rightarrow V(H)$ is a {\em homomorphism
of $G$ to $H$} if $f(u)f(v)$ is an arc of $H$ whenever $uv$ is an arc of $G$ \cite{hombook}.
The {\em product} $G \times H$ of digraphs $G$ and $H$ has the vertex set $V(G) \times V(H)$
and there is an arc in $G \times H$ from $(u,v)$ to $(u',v')$ if $G$ has an arc from $u$ to $u'$
and $H$ has an arc from $v$ to $v'$. The {\em power} $H^k$ is recursively defined as
$H^1=H$ and $H^{k+1}=H \times H^k$. A {\em polymorphism} of $H$ is a homomorphism
$f : H^k \rightarrow H$, for some positive integer $k$. Polymorphisms are of interest in
the solution of constraint satisfaction problems \cite{cohen,jeav}. We say that
polymorphisms $f, g : H^2 \rightarrow H$ are {\em lattice polymorphisms} of $H$, if each
$f$ and $g$ is an associative, commutative, and idempotent, and if moreover $f$ and $g$
satisfy the absorption identities $f(u,g(u,v))=g(u,f(u,v))=u$. It is easy to see that the usual
operations of minimum $f(u,v)=\min(u,v)$ and maximum $g(u,v)=\max(u,v)$, with respect
to a fixed linear ordering $<$, are polymorphisms if and only if $<$ is a Min-Max ordering.
It is also clear that they satisfy the lattice axioms. Thus a digraph which has a Min-Max
ordering does admit lattice polymorphisms. In fact, a digraph admits a Min-Max ordering
if and only if it admits lattice polymorphisms $f, g$ that are {\em conservative}, i.e., satisfy
$f(u,v) \in \{u,v\}, g(u,v) \in \{u,v\}$. (To see that conservative lattice polymorphisms $f, g$
yield a Min-Max ordering, note first that for $u \neq v$ we must have $f(u,v) \neq g(u,v)$
because of the absorption identities, and then let $u < v$ whenever $f(u,v)=u, g(u,v)=v$:
associative and commutative laws imply transitivity of $<$, whence $<$ is a Min-Max
ordering.) Thus we are describing a forbidden structure characterization (and a polynomial
time recognition algorithm) of digraphs with conservative lattice polymorphisms.

An undirected graph (viewed as a symmetric digraph) admits a Min-Max ordering if and
only if each component is either a reflexive proper interval graph or an irreflexive proper
interval bigraph \cite{mincostungraph}. Thus digraphs with Min-Max orderings can be
viewed as digraph analogues of proper interval graphs.
In some cases, we can also describe a geometric representation of digraphs with
Min-Max orderings. A {\em proper adjusted interval digraph} is a digraph $H$ such
that there exist a family of interval pairs $I_v, J_v, v \in V(H)$, where each pair
$I_v, J_v$ share the same left endpoint, no $I_v$ contains another $I_w, w \neq v$,
no $J_v$ contains another $J_w, w \neq v$, and $uv \in E(H)$ if and only if
$I_u \cap J_v \neq \emptyset$. It is not difficult to check that a reflexive digraph
is a proper adjusted interval digraph if and only if it admits a Min-Max ordering.

Proper interval graphs (and bigraphs) are characterized by simple forbidden
structures, and recognized in polynomial time \cite{spin}.
In this paper, we give a polynomial characterization of
digraphs with a Min-Max ordering, suggesting that these digraph
analogues also have interesting structure. Our characterization is in terms of
a novel forbidden structure, which we call a symmetrically invertible pair.
We call our characterization `duality' in the broad sense of having
the presence of some structure (Min-Max ordering) certified by the
absence of some other (forbidden) structure.

We give a similar characterization of digraphs with certain extended
Min-Max orderings, of interest in minimum cost homomorphism problems.
The {\em minimum cost homomorphism problem} for $H$, denoted MinHOM($H$),
asks whether or not an input digraph $G$, with integer costs $c_i(u)$, $u\in V(G)$,
$i \in V(H)$, and an integer $k$, admits a homomorphism to $H$ of total cost
$\sum_{u\in V(G)}c_{f(u)}(u)$ not exceeding $k$. The problem MinHOM($H$)
was first formulated in \cite{gutinDAMlora}; it unifies and generalizes several
other problems \cite{halld2001,jansenJA34,jiangGT32,kroon1997,supowitCAD6},
including two other well studied homomorphism problems, the problem HOM$(H)$
asking for just the existence of homomorphisms \cite{hellJCT48}, and the problem
ListHOM($H$) asking for the existence of homomorphisms in which vertices of $G$
map to vertices of $H$ on allowed lists \cite{pavol}. For undirected graphs $H$, the
complexity of both problems has been classified \cite{hellJCT48,pavol}, and so has
the complexity of the problem MinHOM$(H)$ \cite{mincostungraph}. In each case,
the classification is a dichotomy, in the sense that each problem HOM$(H)$ is
polynomial time solvable or NP-complete. For digraphs, the dichotomy of HOM$(H)$
is an important unproved conjecture, equivalent to the so-called CSP Dichotomy
Conjecture \cite{fv,kroch}. Recent progress specifically on classifying the complexity
of HOM($H$) for classes of digraphs $H$ was reported in \cite{barto1,barto2}.
A simple dichotomy classification of ListHOM($H$) for reflexive digraphs
is described in \cite{lists}; for general digraphs dichotomy follows from
more the general results in \cite{bula}. A simple dichotomy classification of
MinHOM$(H)$ for reflexive digraphs can be found in \cite{arv}. It follows from
\cite{mincostungraph,arv} that both for symmetric digraphs (undirected graphs)
and for reflexive digraphs, MinHOM$(H)$ is polynomial time solvable if $H$ admits
a Min-Max ordering, and is NP-complete otherwise. This is not the case for general
digraphs, as certain extended Min-Max orderings (defined in a later section) also
imply a polynomial time algorithm \cite{yeo}. However, it was conjectured in
\cite{yeo} that MinHOM$(H)$ is NP-complete unless $H$ admits an extended
Min-Max ordering. Several special cases of the conjecture have been verified
\cite{arv,mincostungraph,gutinDAM,yeo,arash}.
We apply our characterization of digraphs with extended Min-Max
ordering to prove this conjecture, obtaining a simple dichotomy classification of the
minimum cost homomorphism problems in digraphs.

As can be expected, one can define minimum cost homomorphism problems for
homomorphisms of more general relational structures $H$ (instead of just one
binary relation, $H$ may have a finite number of finitary relations). In \cite{jonsson},
the authors define, for each relational structure $H$, such a minimum cost constraint
satisfaction problem MinCSP($H$). Even more generally, in \cite{cohenJAIR22}, the
authors define `soft' constraint satisfaction problems, where each hard constraint
(of preserving a $k$-ary relation) is replaced by a cost function assigning
a cost to mapping any $k$-tuple to any other $k$-tuple. Thus MinCSP($H$)
problems can be thought of as having soft unary constraints, with the other
constraints being `hard'. Our results can be directly extended to relational
structures $H$ containing any number of binary relations. On the other
hand, it follows from work of A. Bulatov (personal communication)
that if dichotomy of MinHOM holds for structures with binary relations,
then it holds for all structures. Another proof of dichotomy of MinCSP problems
(but not of our simple classification) has recently been announced in \cite{archiv}.

\section{Min-Max Orderings}

If $uv \in E(H)$, we say that $uv$ is an arc of $H$, or that $uv$ is a {\em forward arc} of
$H$; we also say that $vu$ is a {\em backward arc} of $H$. In any event, we say that $u, v$
are {\em adjacent} in $H$ if $uv$ is a forward or a backward arc of $H$ (and we often use
{\em arc} in this more general sense). The {\em net length}
of a walk is the number of forward arcs minus the number of backward arcs. (Note that a walk
has a designated first and last vertex. For a closed walk we may always choose a direction in
which the net length is non-negative.) An {\em oriented walk} is a walk in which each consecutive
arc is {\em either} a forward arc {\em or} a backward arc. A digraph is {\em balanced} if it does not
contain an oriented cycle of non-zero net length. It is easy to see that a digraph is balanced if
and only if it admits a {\em labeling} of vertices by non-negative integers so that each arc goes
from some level $i$ to the level $i+1$. The {\em height} of $H$ is the maximum net length
of a walk in $H$. Note that an unbalanced digraph has infinite height, and the height of a balanced
digraph is the greatest label in a non-negative labeling in which some vertex has label zero.

For any walk $P = x_0, x_1, \dots, x_n$ in $H$, we consider the {\em minimum height} of $P$ to
be the smallest net length of an initial subwalk $x_0, x_1, \dots, x_i$, and the {\em maximum height}
of $P$ to be the greatest net length of an initial subwalk $x_0, x_1, \dots, x_i$. Note that when $i=0$,
we obtain the trivial subwalk $x_0$ of net length zero, and when $i=n$, we obtain the entire walk $P$.
We shall say that $P$ is {\em constricted from below} if the minimum height of $P$ is zero (no initial
subwalk $x_0, x_1, \dots, x_i$ has negative net length), and {\em constricted} if moreover the
maximum height is the net length of $P$ (no initial subwalk $x_0, x_1, \dots, x_i$ has greater net
length than $x_0, x_1, \dots, x_n$). We also say that $P$ is {\em nearly constricted from below}
if the net length of $P$ is minus one, but all proper initial subwalks $x_0, x_1, \dots, x_i$ with
$i < n$ have non-negative net length. It is easy to see that a walk which is nearly constricted from
below can be partitioned into two constricted pieces, by dividing it at any vertex achieving the
maximum height.

A vertex $x$ of $H$ is called {\em extremal} if every walk starting in $x$ is constricted from
below, i.e., there is no walk starting in $x$ with negative net length. It is clear that a balanced
digraph $H$ contains extremal vertices (we can take any $x$ from which starts a walk with
net length equal to the height of $H$), and an unbalanced digraph does not have extremal
vertices (from any $x$ we can find a walk of negative net length by going to an unbalanced
cycle and then following it long enough in the negative direction). Moreover, in a weakly
connected digraph $H$, any extremal vertex $x$ is the beginning of a constricted walk
of net length equal to the height of $H$.

For walks $P$ from $a$ to $b$, and $Q$ from $b$ to $c$, we denote by $PQ$ the walk from
$a$ to $c$ which is the concatenation of $P$ and $Q$, and by $P^{-1}$ the walk $P$ traversed
in the opposite direction, from $b$ to $a$. We call $P^{-1}$ the {\em reverse} of $P$. For a
closed walk $C$, we denote by $C^a$ the concatenation of $C$ with itself $a$ times.

Our main result is the following forbidden structure characterization.

\begin{theorem}\label{main}
A digraph $H$ admits a Min-Max ordering if and only if it does not contain an induced
unbalanced oriented cycle of net length greater than one, and does not contain a
symmetrically invertible pair.
\end{theorem}

An oriented cycle of $H$ is {\em induced} if $H$ contains no other arcs on the vertices of the
cycle. In particular, an induced oriented cycle of length greater than one does not contain a
loop. Symmetrically invertible pairs are defined below.

We define two walks $P = x_0, x_1, \dots, x_n$ and $Q = y_0, y_1, \dots, y_n$ in $H$ to be
{\em congruent}, if they follow the same pattern of forward and backward arcs, i.e., $x_ix_{i+1}$
is a forward (backward) arc if and only if $y_iy_{i+1}$ is a forward (backward) arc (respectively).
Suppose the walks $P, Q$ as above are congruent. We say an arc $x_iy_{i+1}$ is {\em a faithful
arc from $P$ to $Q$}, if it is a forward (backward) arc when $x_ix_{i+1}$ is a forward (backward)
arc (respectively), and we say an arc $y_ix_{i+1}$ is {\em a faithful arc from $Q$ to $P$}, if it is
a forward (backward) arc when $x_ix_{i+1}$ is a forward (backward) arc (respectively). We say
that $P, Q$ {\em avoid each other} if there is no pair of faithful arcs $x_iy_{i+1}$ from $P$ to $Q$,
and $y_ix_{i+1}$ from $Q$ to $P$, for some $i=0, 1, \dots, n$. A {\em symmetrically invertible pair},
or {\em sym-invertible pair}, in $H$ is a pair of distinct vertices $u, v$, such that there exist
congruent walks $P$ from $u$ to $v$ and $Q$ from $v$ to $u$, which avoid each other.

A somewhat different notion of invertible pairs occurs in the study of list homomorphisms \cite{lists},
and so we add the adjective `symmetrically' or the prefix `sym-' to distinguish the two concepts.

We define an auxiliary digraph $H^*$ as follows. The vertices of $H^*$ are all ordered pairs $(x,y)$
of distinct vertices of $H$, and there is an arc in $H^*$ from $(x,y)$ to $(x',y')$ just if $xx', yy'$ are
both forward arcs of $H$ but $xy', yx'$ are not both forward arcs of $H$. (Either just one is an arc,
or neither is an arc). Note that in $H^*$ we have an arc from $(x,y)$ to $(x',y')$ if and only if there
is an arc from $(y,x)$ to $(y',x')$. It follows from these definitions that a sym-invertible pair of vertices
$u, v$ in $H$ corresponds in $H^*$ to an oriented path between vertices $(u,v)$ and $(v,u)$, i.e.,
$H$ admits a sym-invertible pair if and only if there exist $u, v$ so that $(u,v)$ and $(v,u)$ belong
to the same weak component of $H^*$.

\begin{theorem}\label{inv}
If $H$ contains a sym-invertible pair, then it does not admit a Min-Max ordering.
\end{theorem}

\pf Indeed, suppose that $<$ is a Min-Max ordering and $(x,y)$ and $(x',y')$ are adjacent in $H^*$. Observe
that if $x$ precedes (respectively follows) $x'$ in $<$, then $y$ must also precede (respectively follow) $y'$
in $<$. Hence if $u, v$ is a sym-invertible pair in $H^*$, then if $u$ is ordered before (respectively after)
$v$, by following the avoiding congruent walks $P$ and $Q$ from the definition of a sym-invertible pair,
we conclude that also $v$ must be ordered before (respectively after) $u$. So, a sym-invertible pair
implies a violation of antisymmetry, and hence it is an obstruction to the existence of a Min-Max ordering.
\qed

\begin{theorem}\label{cyc}
If $H$ contains an induced unbalanced oriented cycle of net length greater than one,
then it does not admit a Min-Max ordering.
\end{theorem}

\pf
Indeed, suppose $C$ is an induced unbalanced oriented cycle of net length $k>1$, and let $x_0$
be a vertex of $C$ in which we can start a walk $P$ around $C$ which is constricted from below.
It is easy to see that such a vertex must exist; in fact, we may assume that even $P \setminus x_0$
is constricted from below. Then following $P$ let $x_i$ ($1 \leq i \leq k-1$) be the last vertex on $P$
such that the walk from $x_0$ to $x_i$ has net length $i$. It is easy to see that $x_i, i=0,1,\dots,k-1$
are all found in the first pass around $C$. Then $(x_0,x_1), (x_1,x_2), \dots, (x_{k-1},x_0)$ belong to
the same weak component of $H^*$, in violation of transitivity of $<$. Indeed, it is easy to prove, using
Lemma \ref{typer} and the fact that $C$ is an induced cycle, that any two pairs $(x_{i-1},x_i)$ and
$(x_i,x_{i+1})$ belong to the same weak component of $H^*$.
\qed

Thus we shall assume that the digraph $H$ has no induced unbalanced cycle of net length greater
than one, and no sym-invertible pair.

We shall frequently use the following key lemma.

\begin{lemma}\label{core}
Let $a, b, c$ be three vertices of $H$, such that the component
of $H^*$ which contains $(a,b)$ contains neither of $(a,c), (c,b)$.

Let $A, B, C$ be congruent walks starting at $a, b, c$ respectively.

If $A$ and $B$  avoid each other, then $B$ and $C$ also avoid each
other, and $A$ and $C$ also avoid each other.
\end{lemma}

\pf By symmetry, it suffices to prove the claim about $B$ and $C$.

Suppose $A=a_1,a_2,\dots,a_n$, $B=b_1,b_2,\dots,b_n$, and
$C=c_1,c_2,\dots,c_n$ (here $a_1=a$, $b_1=b$, and $c_1=c$).
For a contradiction, suppose that $B$ and $C$ do not avoid each
other, and let $i$ be the least subscript such that both $b_ic_{i+1}$
and $c_ib_{i+1}$ are faithful arcs in $H$. (Note that $i$ could be
equal to $n-1$.)

Since $(a,b)$ and $(a,c)$ are not in the same component of $H^*$,
the congruent walks $$R = a_1,\dots,a_i,a_{i+1},a_i,\dots,a_1
{\rm \ \ and \ \ }
S = b_1,\dots,b_i,b_{i+1},c_i,\dots,c_1$$ do not  avoid each other.
Since $A$ and $B$ do avoid each other, any faithful arcs between
$R$ and $S$ must be between $b_{i+1},c_i,\dots,c_1$ and
$a_{i+1},a_i,\dots,a_1$. Suppose first there exists a subscript
$j<i$ such that $a_jc_{j+1}$ and $c_ja_{j+1}$ are faithful arcs,
and let $j$ to be chosen as small as possible subject to this.
Note that there is a second possibility, that $a_ib_{i+1}$ and
$c_ia_{i+1}$ are the only faithful arcs. We think of this case as
having $j=i$, with the understanding that $c_{j+1}$ is replaced
by $b_{j+1}$, and we will deal with it at the end of this proof.

Since $(a,b)$ and $(c,b)$ are not in the same component of $H^*$,
the congruent walks $$R'=a_1,\dots,a_j,a_{j+1},c_j,\dots,c_1 {\rm \ \ and \ \ }
S'=b_1,\dots,b_j,b_{j+1},b_j,\dots,b_1$$ do not avoid each other.
Since $A$ and $B$ do avoid each other and since $j < i$ while
$i$ was chosen to be minimal, the faithful arcs must be
$b_ja_{j+1},c_jb_{j+1}$. Similarly, the congruent walks $$R'' = a_1,\dots,a_j,
c_{j+1},c_j,\dots,c_1 {\rm  \ \ and \ \ } S'' = b_1,\dots,b_j,b_{j+1},b_j,
\dots,b_1$$ yield the faithful arcs $a_jb_{j+1}$ and $b_jc_{j+1}$ -
contradicting the fact that $A, B$  avoid each other.

Returning now to the special case when $j=i$, we observe that we
can use the same pair of walks $R', S'$ as above and then modify
the walks $$R'' = a_1,\dots,a_i,a_{i+1},c_i,\dots,c_1 {\rm \ \ and \ \ }
S'' = b_1,\dots,b_i,c_{i+1},b_i,\dots,b_1,$$ to conclude that
$b_ia_{i+1}$ is again an arc, yielding the same contradiction.
\qed

We note that two congruent paths which avoid each other can not
intersect, thus the lemma implies that $B$ and $T$ are disjoint.

The following lemma is well known. (For a proof, see \cite{don,zhu}
or Lemma 2.36 in \cite{hombook}).

\begin{lemma}\label{typer}
Let $P_1$ and $P_2$ be two constricted walks of net length $r$.
Then there is a constricted path $P$ of net length $r$ that admits
a homomorphism $f_1$ to $P_1$ and a homomorphism $f_2$ to
$P_2$, such that each $f_i$ takes the starting vertex of $P$ to the
starting vertex of $P_i$ and the ending vertex of $P$ to the
ending vertex of $P_i$.
\end{lemma}

We shall call $Q$ a {\em common pre-image} of $P_1$ and $P_2$.

We now formulate a corollary of the last two lemmas which will be
used frequently.

\begin{corollary}\label{pes}
Let $a, b, c$ be three vertices of $H$, such that the component
of $H^*$ which contains $(a,b)$ contains neither of $(a,c), (c,b)$.

Let $A, B, C$ be three constricted walks of the same net length,
starting at $a, b, c$ respectively. Suppose that $A$ and $B$ are
congruent and avoid each other.

Then there exists congruent common pre-images $A', B', C'$ of
$A, B, C$ starting at $a, b, c$ respectively, such that $B'$ and
$C'$ avoid each other, and $A'$ and $C'$ also avoid each other.
\end{corollary}

We note that Corollary \ref{pes} will sometimes be applied to walks
that are not constricted but can be partitioned into constricted
walks of corresponding net lengths.

Since $H$ has no sym-invertible pairs, we conclude that if a pair $(u,v)$
is in a weak component $C$ of $H^*$, then the corresponding reversed
pair $(v,u)$ is in a different component $C' \neq C$ of $H^*$. Moreover,
if any $(x,y)$ also lies in $C$, then the corresponding reversed $(y,x)$
must also lie in $C'$, since reversing all pairs on an oriented walk between
$(u,v)$ and $(x,y)$ results in an oriented walk between $(v,u)$ and $(y,x)$.
Thus the components of $H^*$ come in pairs $C, C'$ so that the ordered
pairs in $C'$ are the reverses of the ordered pairs in $C$. We say the
components $C, C'$ are {\em dual} to each other.

\section{The Algorithm}

We assume that $H$ has no induced unbalanced cycle of
net length greater than one, and no sym-invertible pairs.
We shall give an algorithm to construct a desired Min-Max ordering $<$.
At each stage of the algorithm,
some components of $H^*$ have already been {\em chosen}. The chosen
components define a binary relation $<$ as follows: we set $a < b$
if the pair $(a,b)$ belongs to one of the chosen components. Whenever
a component $C$ of $H^*$ is chosen, its dual component $C'$ is {\em
discarded}. The objective is to avoid a {\em circular chain}
$$(a_0,a_1), (a_1,a_2), \dots, (a_n,a_0)$$
of pairs belonging to the chosen components. Our algorithm always
chooses a component $X$ of maximum height from among the as yet
un-chosen and un-discarded components. If $X$ creates a circular chain,
then the algorithm chooses the dual component $X'$. We shall show
that at least one of $X$ and $X'$ will not create circular chain. (Note that
this implies that the component $X$ does not contain a circular chain.)
Thus at the end of the algorithm we have no circular chain and hence $<$
is a total order. It is easy to see that $<$ is a Min-Max ordering. Indeed,
if $i<j, s<r$ and $ir, js \in A(H)$ but $is \not\in A(H)$ or $jr \not\in A(H)$,
then $(i,j)$ and $(r,s)$ are adjacent in $H^*$ - whence we have either
$i<j, r<s$ or $j<i, s<r$, contrary to what was supposed.

\begin{theorem}\label{sunshine}
The algorithm avoids creating a circular chain.
\end{theorem}

Thus suppose that at a certain time $T$ there was no circular chain amongst
the chosen components, that $X$ had the maximum height from all unchosen
(and undiscarded) components, and that the addition of $X$ to the chosen
components created the circular chain $(a_0,a_1), (a_1,a_2), \dots, (a_n,a_0)$,
and the addition of the dual component $X'$ created the circular chain
$(b_0,b_1), (b_1,b_2), \dots, (b_m,b_0)$. We may suppose that $T$ was
minimum for which this occurs, then $n$ was minimum value for this $T$,
and then $m$ was minimum value for this $T$ and $n$. We may also assume
that $X$ contains the pairs $(a_n,a_0), (b_0,b_m)$, and possibly other
$(a_{i},a_{i+1})$ or $(b_j,b_{j+1})$.

Let $A_i$ be the weak component of $H^*$ containing the pair $(a_i,a_{i+1})$,
and $B_j$ be the weak component containing the pair $(b_j,b_{j+1})$;
subscripts are modulo $n$ and $m$ respectively. (Thus $X=A_n=B_m'$.)
Note that the minimality of $n$ implies that no $A_i$ contains a pair
$(a_k,a_{\ell})$ for subscripts (reduced modulo $n+1$) $\ell \neq k+1$
(and similarly for $B_j$). (This is helpful when checking the hypothesis
of Lemma \ref{core} and Corollary \ref{pes}, as in Case 2 below.)

The following lemma is our basic tool.

\begin{lemma}\label{itit}
Suppose that none of the pairs $(a_i,a_{i+1})$ is extremal in its component $A_i$.

Then there exists another circular chain $(a'_0,a'_1), (a'_1,a'_2), \dots, (a'_n,a'_0)$
where each $(a'_i,a'_{i+1})$ can be reached from the corresponding $(a_i,a_{i+1})$
by a walk in $A_i$ nearly constricted from below.
\end{lemma}

\pf
Since $(a_i,a_{i+1})$ is not extremal, there exists a walk $W_i$ in $A_i$ from
$(a_i,a_{i+1})$ to some $(p_i,q_i)$, which is nearly constricted from below.
Corresponding to this walk in $A_i$, there are two walks $P_i$ and $Q_i$ in $H$,
from $a_i$ to $p_i$ and from $a_{i+1}$ to $q_i$ respectively, which avoid each
other. Let $L_i$ be the maximum height of $W_i$ (which is the same as in $P_i$,
and $Q_i$).

We now explain how to choose $n$ of the $2n$ vertices $p_i, q_i$ which also form a
circular chain. For any $i$, instead of $a_i$, we choose $a'_i=q_{i-1}$ if $L_{i-1} < L_i$,
and we choose $a'_i=p_i$ otherwise. We now show that $(a'_0,a'_1), (a'_1,a'_2), \dots,
(a'_n,a'_0)$ is a circular chain; it suffices to show that each $(a'_i,a'_{i+1})$ is in $A_i$.

\noindent
{\bf Case 1.} Suppose $L_i \leq L_{i-1}$ and $L_i \leq L_{i+1}$.

In this case, we have $a'_i=p_i, a'_{i+1}=q_i$, and $(p_i,q_i)$ is
in $A_i$ by definition.

\noindent
{\bf Case 2.} Suppose $L_i \geq L_{i-1}$ and $L_i \geq L_{i+1}$.

In this case, we have $a'_i=q_{i-1}, a'_{i+1}=p_{i+1}$. We may assume that
$L_{i+1} \leq L_{i-1}$ (otherwise the argument is symmetric). Consider the
congruent walks $A = P_{i-1}$ from $a_{i-1}$ to $p_{i-1}$ and $B = Q_{i-1}$
from $a_i$ to $q_{i-1}$. They are nearly constricted from below, and have
maximum height $L_{i-1}$. Consider the following walk $C$ from $a_{i+1}$
to $p_{i+1}$: the walk $C$ starts with a portion of $Q_i$, up to the maximum
height $L_{i-1}$ and then back down to $a_{i+1}$, followed by $P_{i+1}$.
Note that $C$ is also nearly constricted from below and has the same maximum
height $L_{i-1}$. It follows that $A, B, C$ can each be partitioned into two
constricted pieces of corresponding net lengths. Since
$(a_{i-1},a_{i+1}), (a_{i+1},a_i) \not\in A_{i-1}$ by the minimality of $n$,
Corollary \ref{pes} (applied to each of the constricted pieces) implies that
$B$ and $C$ avoid each other. Since $a'_i=q_{i-1}, a'_{i+1}=p_{i+1}$, we
have a walk in $H^{*}$ from $(a_i,a_{i+1})$ to $(a'_i,a'_{i+1})$, whence
$(a'_i,a'_{i+1}) \in A_i$.

\noindent
{\bf Case 3.} Suppose $L_{i-1} < L_i < L_{i+1}$
(or $L_{i-1} > L_i > L_{i+1}$).

In this case, we have $a'_i=q_{i-1}, a'_{i+1}=q_i$. Since the subscripts are
computed modulo $n+1$, there must exist a subscript
$s$ such that $L_s \geq L_i \geq L_{s+1}$. Now we again apply Corollary \ref{pes}
to the walks $A=P_i, B=Q_i$, and $C$ from $a_{s+1}$ to $p_{s+1}$ using $P_{s+1}$
and a portion of $Q_s$, to conclude that $C$ avoids $B$. Finally, we once more apply
Lemma \ref{core} to the three walks $B, C$, and $D$ from $a_i$ to $a'_i=q_{i-1}$
using $Q_{i-1}$ and a portion of $P_i$, to conclude that $D$ avoids $B$. Hence
there is a walk in $H^{*}$ from $(a_i,a_{i+1})$ to $(a'_i,a'_{i+1})$, implying that
$(a'_i,a'_{i+1}) \in A_i$.
\qed

We now continue with the {\bf proof of Theorem \ref{sunshine}}.

We distinguish two principal cases, depending on
whether or not the component $X$ is balanced.

{\bf We first assume that the component $X$ is balanced.}

Suppose the height of $X$ is $h$.

\begin{lemma}\label{cuc}
Suppose some $(a_k,a_{k+1})$ is extremal in $A_k$.

Let $(a_i,a_{i+1}), (a_j,a_{j+1})$ be distinct non-extremal pairs in $A_i, A_j$
respectively, and let $W_i, W_j$ be walks in $A_i, A_j$ starting from
$(a_i,a_{i+1}), (a_j,a_{j+1})$ respectively, that are nearly constricted from
below. Let $L_i, L_j$ be the maximum heights of $W_i, W_j$ respectively.

Then $L_i > h$ or $L_j > h$.
\end{lemma}

\pf
Suppose $L_i \leq h, L_j \leq h$, and assume, without loss of generality, that
$L_i \leq L_j$. Since some $(a_k,a_{k+1})$ is extremal, we may assume that
neither $(a_{i-1},a_i)$, nor $(a_{j+1},a_{j+2})$ initiate walks of negative net
length with maximum height at most $h$. Thus each of $(a_{i-1},a_i)$,
$(a_{j+1},a_{j+2})$ is either extremal, and thus initiate a constricted walk
of net length $h$, or initiates a walk of negative net length, with maximum
height greater than $h$, and hence again initiates a constricted walk of net
length $h$. Thus we have

\begin{itemize}
\item
a constricted walk $U_i$ of net length $h$ from $a_i$
\item
a walk $V_i$, nearly constricted from below, from $a_i$ to some $p$
\item
a constricted walk $U_{j+1}$ of net length $h$ from $a_{j+1}$
\item
a constricted walk $U_{j+2}$ of net length $h$ from $a_{j+2}$, which avoids
$U_{j+1}$ and is congruent to it
\item
a walk $V_j$, nearly constricted from below, from $a_j$, and
\item
a walk $V_{j+1}$, nearly constricted from below, from $a_{j+1})$ to some $q$,
which avoids $V_j$ and is congruent to it.
\end{itemize}

Consider the three walks $A, B, C$, where $A$ is the reverse of $V_{j+1}$
(starting in $q$), $B$ is the reverse of $V_j$, and $C$ is the reverse of $V_i$
followed by a suitable piece of $U_i$ (and its reverse) as needed to have the
same maximum height $L_j$ as $V_j$. Each of these walks consists of two
constricted pieces and hence we can apply Corollary \ref{pes} twice to conclude
that there exist congruent pre-images $A'$ and $C'$ of $A$ and $C$ respectively,
which avoid each other. We can also apply Corollary \ref{pes} to the constricted
walks $U_{j+1}, U_{j+2}, U_i$ to conclude that there are congruent pre-images
$A'', C''$ of $U_{j+1}, U_j$ respectively, which avoid each other. Concatenating
$A'$ with $A''$ and $C'$ with $C''$, we conclude that $(p,q)$ belongs to a
component of $H^*$ which has height greater than $h$; this means that before
$X$ we should have chosen the component of $H^*$ containing $(a_i,a_j)$,
which is a contradiction.
\qed

\begin{lemma}\label{ext}
If any $(a_i,a_{i+1})$ is extremal in $A_i$, then $(a_n,a_0)$ is extremal in $X=A_n$.
\end{lemma}

\pf
Suppose $(a_n,a_0)$ is not extremal. By Lemma \ref{cuc}, it remains to
consider the case when both $(a_0,a_1)$ and $(a_{n-1},a_n)$ are extremal.
Since $(a_0,a_1)$ is extremal, there exists a constricted walk in $H^*$ starting
from $(a_0,a_1)$ of net length equal to the height of $A_0$, which is at least
$h$, according to our algorithm. Similarly, there exists a constricted walk from
$(a_{n-1},a_n)$ of net length equal to the height of $A_{n-1}$, which is also
at least $h$. From the walk in $A_{n-1}$, we extract a constricted walk $A$
starting in $a_{n-1}$, and a congruent constricted walk $B$ starting in $a_n$
such that $A, B$ have net length $h$ and avoid each other. From the walk in
$A_0$ we moreover extract a walk $C$ starting in $a_0$ which is also constricted
and has net length $h$. Now Corollary \ref{pes} ensures that $B$ and $C$ have
congruent pre-images $B'$ and $C'$ which avoid each other. Let $B'', C''$ be
two congruent walks of negative net length from $a_n, a_0$ respectively, which
avoid each other; such walks exist since $(a_n,a_0)$ is not extremal. Now taking
the concatenations of $(B'')^{-1}$ with $B'$ and $(C'')^{-1}$ with $C'$ yields a
walk in $X$ of net length greater than $h$, which is a contradiction.
\qed

Thus Lemma \ref{itit} ensures that we may assume that $(a_n,a_0)$ is extremal
in $X$ (and similarly for $(b_0,b_m)$). The proof now distinguishes whether or
not $X$ contains another pair $(a_i,a_{i+1})$ (or similarly for $(b_j,b_{j+1})$).

Suppose first that some $(a_i,a_{i+1}) \in X$, and let $W$ be a walk from
$(a_n,a_0)$ to $(a_i,a_{i+1})$ in $X$. We observe that the net length
of $W$ must be zero. Indeed, since $(a_n,a_0)$ is extremal in $X$, the
net length of $W$ must be non-negative. If the net length were positive,
then $W^{-1}$ would be a walk from $(a_i,a_{i+1})$ of negative net length
and with maximum height less than $h$. Thus Lemma \ref{cuc} implies that
both $(a_{i-1},a_i), (a_{i+1},a_{i+2})$ initiate walks of net length $h$,
yielding walks $U_{i-1}, U_i, U_{i+1}, U_{i+2}$ of net length $h$, from
$U_{i-1}, U_i, U_{i+1}, U_{i+2}$, respectively. Here $U_{i-1}, U_i$ are
congruent constricted walks that avoid each other, and hence Corollary
\ref{pes} implies that there are pre-images of $U_i, U_{i+1}$ of net length
$h$ that are congruent and avoid each other. This yields a walk in $X$
from $(a_i,a_{i+1})$ of net length $h$ - and concatenated with $W$ we
obtain a walk in $X$ from $(a_n,a_0)$ of net length strictly greater than
$h$, which is impossible.

Thus the net length of $W$ is zero, and hence it can be partitioned into
two constricted pieces, $U$ from $(a_n,a_0)$ to some vertex $(z_1,z_2)$
of maximum height, and $V$ from $(z_1,z_2)$ to $(a_i,a_{i+1})$. Let
$U_1$ (respectively $U_2$) denote the corresponding walk from $a_n$
to $z_1$ (respectively from $a_0$ to $z_2$), and similarly for $V_1, V_2$.
Then Lemma \ref{core} applied to $U_1, U_2, V_2$ implies that $(z_1,z_2)$
and $(a_n,a_{i+1})$ are in the same component of $H^*$; however,
$(z_1,z_2) \in X$, so $(a_n,a_{i+1}) \in X$, contrary to the minimality
of $n$.

Thus we conclude that $X$ does not contain another $(a_i,a_{i+1})$ or
$(b_j,b_{j+1})$. In other words, before time $T$ we have the chosen all
the pairs
$$(a_0,a_1), \dots, (a_{n-1},a_n), (b_0,b_1), \dots, (b_{m-1},b_m),$$
and then at time $T$ we chose the component $X$ containing
$(a_n,a_0)$ as well as $(b_0,b_m)$. Consider a fixed walk $W$
in $X$ from $(a_n,a_0)$ to $(b_0,b_m)$. Since $(a_n,a_0)$, and
by symmetry also $(b_0,b_m)$, is extremal, $W$ must have net
length zero. Moreover, we may assume that $W$ reaches some
vertex $(z_1,z_2)$ of maximum height $h$. Thus $W$ consists
of two constricted walks $U, V$. Let again $U_1$ (respectively
$U_2$) be the corresponding walk in $H$ from $a_n$ (respectively
from $a_0$) to a vertex of maximum height, and similarly let $V_1$
(respectively $V_2$) be the corresponding walks from the vertices
of maximum height to $b_0$ (respectively $b_m$).

We shall prove first that there is a constricted walk  of net length
$h$ from $a_1$. Indeed, the component $A_0$, containing the
vertex $(a_0,a_1)$ must have height at least $h$, according to
the rules of our algorithm. If $(a_0,a_1)$ does not initiate a walk
of net length $h$, it must not be extremal, i.e., it must initiate a
walk of negative net length. The same argument yields a walk of
negative net length from $(a_1,a_2)$. Since such walks contain
walks that are nearly constricted from below, we obtain a contradiction
with Lemma \ref{cuc}. A similar argument applies to $b_1$.

Thus there are constricted walks of net length $h$ from both $a_1$
and $b_1$, say $R$ and $S$ respectively. We can now use Corollary
\ref{pes} on the walks $A=U_1, B=U_2, C=R$, and again on the walks
$A=V_1, B=V_2, C=R^{-1}$ to deduce that $U_2$ concatenated with
$V_2$ and $R$ concatenated with $R^{-1}$ avoid each other, whence
$(a_0,a_1)$ and $(b_m,a_1)$ are in the same component of $H^*$.
By a similar argument we also deduce that $(b_0,b_1)$ and $(a_n,b_1)$
are also in the same component of $H^*$. This is impossible, as it would
mean that at time $T-1$ there already was a circular chain, namely
$(b_m,a_1),(a_1,a_2), \dots, (a_n,b_1), (b_1,b_2), \dots, (b_{m-1},b_m)$.

This completes the proof of Theorem \ref{sunshine} in case $X$ is balanced.

{\bf We now assume the component $X$ is unbalanced.}

In this case, the rules of the algorithm imply that each component $A_i$
and $B_j$ is also unbalanced. Thus each of the components contains an
oriented cycle of net length one, and hence there is a closed walk of net
length one, or minus one, starting in any vertex in any of these components.
In particular, as we observed before, an unbalanced digraph does not contain
any extremal vertices. We shall define a vertex $u$ in an unbalanced digraph
to be {\em weakly extremal} if there is a walk starting from $u$ which is
constrained from below and has infinite maximum height. Each oriented
cycle of positive net length, and hence each unbalanced digraph, contains
a weakly extremal vertex.

Recall our assumptions that $X$ contains $(a_n,a_0), (b_0,b_m)$ and maybe
other pairs, creating the circular chain $(a_0,a_1), (a_1,a_2), \dots, (a_n,a_0)$
in $X$ and the circular chain $(b_0,b_1),$ $(b_1,b_2),$
$\dots, (b_m,b_0)$ in $X'$.

We first claim that we may assume that each $(a_i,a_{i+1})$ (and similarly each
$(b_j,b_{j+1})$) is weakly extremal. Indeed, suppose there is a walk in $A_i$
from $(a_i,a_{i+1})$ to some weakly extremal vertex $(e_i,e_{i+1})$, of negative
net length $\ell_i$. Let $\ell$ be the minimum of all $\ell_i, i=0, \dots, n$. Since
$(a_i,a_{i+1})$ initiates a closed walk in $A_i$ of net length minus one, there is
a walk from each $(a_i,a_{i+1})$ to the weakly extremal vertex $(e_i,e_{i+1})$
of net length $\ell$. Now we apply Lemma \ref{itit} $\ell$ times to obtain a
circular chain $(a'_0,a'_1), (a'_1,a'_2), \dots, (a'_n,a'_0)$. It follows from the
proof of Lemma \ref{itit} that each $(a'_i,a'_{i+1})$ has a walk of net length zero
to $(e_i,e_{i+1})$; this means that each $(a'_i,a'_{i+1})$ is weakly extremal.

As in the balanced case, we first assume that some $(a_i,a_{i+1}) \in X$. Then
there is a walk $W$ from $(a_n,a_0)$ to $(a_i,a_{i+1})$ in $X$ of net length
zero. Indeed, the argument above shows that both $(a_n,a_0)$ and $(a_i,a_{i+1})$
have a walk of net length $\ell$ to $(e_i,e_{i+1})$, since in this case $A_i=A_n=X$.
As before, $X$ can be partitioned into two constricted pieces, $U$ and $V$, and
Lemma \ref{core} implies that $(a_n,a_{i+1}) \in X$, contrary to the minimality
of $n$.

If $X$ does not contain another $(a_i,a_{i+1})$ or $(b_j,b_{j+1})$, we
again proceed as in the balanced case. There exists a walk $W$ in $X$
of net length zero from $(a_n,a_0)$ to $(b_0,b_m)$. (Both $(a_n,a_0)$
and $(b_0,b_m)$ can reach $(e_n,e_0)$ with walks of the same net
length.) Let $L$ be the maximum height of $W$. Thus $W$ consists
of two constricted walks $U, V$. Let again $U_1$ (respectively
$U_2$) be the corresponding walk in $H$ from $a_n$ (respectively
from $a_0$) to a vertex of maximum height, and similarly let $V_1$
(respectively $V_2$) be the corresponding walks from the vertices
of maximum height to $b_0$ (respectively $b_m$). Since $(a_0,a_1)$
is weakly extremal, there is a constricted walk  of net length $L$ from
$a_1$, and for a similar reason, there is such a walk also from $b_1$.

We can now use Corollary \ref{pes} as in the balanced case, to deduce
that $(a_0,a_1)$ and $(b_m,a_1)$ are in the same component of $H^*$,
and that $(b_0,b_1)$ and $(a_n,b_1)$ are in the same component of $H^*$,
yielding the same contradiction.

This completes the proof of Theorem \ref{sunshine}.

\begin{corollary}
The following three statements are equivalent for a digraph $H$

\begin{enumerate}
\item
$H$ admits a Min-Max ordering
\item
$H$ has no invertible pair and no induced oriented cycle of net length
greater than one
\item
no weak component of $H^*$ contains a circular chain
\end{enumerate}

\end{corollary}

\pf
The equivalence of (1) and (2) is Theorem \ref{main}.
It is obvious that (1) implies (3). Finally, (3) implies (2)
as an invertible pair in $H$ is a circular chain of length
two in a weak component of $H^*$, and the proof of
Theorem \ref{cyc} shows that an induced oriented cycle
of net length greater than one yields a circular chain in
a weak component of $H^*$.
\qed

It follows that the existence of a Min-Max ordering can be tested
in polynomial time: to test (2), construct $H^*$, find its weak
components, and test each for circular chains. Testing a weak
component for circular chains amounts to looking at a set of
ordered pairs, i.e., a digraph, and looking for a directed cycle.
Acyclicity can be tested in linear time by topological sort.

\section{Extended Min-Max Orderings}

We now discuss {\em extended Min-Max orderings}, for digraphs $H$
with a fixed homomorphism $\ell : H \rightarrow \vec{C}_k$. For the
remainder of this discussion, the digraph $H$ and the homomorphism
$\ell$ is fixed. (The standard Min-Max orderings may be viewed as the
special case $k=1$.) Assume the vertices of $\vec{C}_k$ are $0, 1, \dots, k-1$,
and let $V_i = \ell^{-1}(i)$. A $k$-{\em Min-Max ordering} of $H$ is a linear
ordering $<$ of each $V_i$, so that the Min-Max condition ($i<j, s<r$
and $ir, js \in A(H)$ imply $is \in A(H)$ and $jr \in A(H)$) is satisfied for
$i, j$ and $s, r$ in any two circularly consecutive sets $V_i$ and $V_{i+1}$
respectively (subscript addition modulo $k$). Note that a Min-Max ordering
is also a $k$-Min-Max ordering for any $k$ and $\ell$; however, there are
digraphs with a $k$-Min-Max ordering that do not have a Min-Max ordering
- for instance $\vec{C}_k$.

We shall consider a subgraph of $H^*$ defined as follows.
The digraph $H^{(k)}$ is the subgraph of $H^*$ induced by all
ordered pairs $(x,y)$ of with $\ell(x)=\ell(y)$. We say that $(u,v)$
is a {\em symmetric $k$-invertible pair} (or a {\em sym-$k$-invertible
pair}) in $H$ if there is in $H^{(k)}$ an oriented walk joining $(u,v)$
and $(v,u)$. Note that each sym-$k$-invertible pair is just a sym-invertible
pair in $H$ in which $u$ and $v$ have $\ell(u)=\ell(v)$. Note that $H$ may
contain sym-invertible pairs, but none with $\ell(u)=\ell(v)$. Consider,
for instance the directed hexagon $\vec{C}_6$ on $0,1,2,3,4,5$.
The pair $0, 3$ is sym-invertible, but not sym-$6$-invertible. Note also
that there is a homomorphism $\ell$ of $\vec{C}_6$ to $\vec{C}_3$
in which $\ell(0)=\ell(3)$, in which the pair $0, 3$ is $3$-invertible.

The extended version of our main theorem is the following.

\begin{theorem}\label{main-k}
A digraph $H$ with a homomorphism $\ell$ to $\vec{C}_k$ admits a
$k$-Min-Max ordering if and only if it does not contain an induced
unbalanced oriented cycle of net length other than $k$, and does
not contain a sym-$k$-invertible pair.
\end{theorem}


\pf
If $H$ contains an induced oriented cycle of net length
$\lambda k$ with $\lambda \neq 1$, then it contains $\lambda$
vertices $a_0, a_1, \dots, a_{\lambda-1}$ and a circular chain
$(a_0,a_1), \dots (a_{\lambda-1},a_0)$ as in the case $k=1$.
If $H$ contains a sym-$k$-invertible pair $a_0, a_1$, then it
contains the circular chain $(a_0,a_1), (a_1,a_0)$.

If $H$ does not contain such a cycle or invertible pair, then the
same algorithm applied to $H^k$ again avoids creating a circular
chain. The proof of this fact is analogous to the case $k=1$. The
only additional twist occurs in the case when $X$ is unbalanced,
where we need to observe that each part $V_i$ must contain a
vertex which is weakly extremal; this is easy to see.
\qed

We apply Theorem \ref{main-k} to prove the following result
conjectured in \cite{yeo}.

\begin{theorem}\label{gut}
If $H$ has a homomorphism to some $\vec{C}_k$ which admits
a $k$-Min-Max ordering then MinHOM($H$) is polynomial time
solvable. Otherwise, MinHOM($H$) is NP-complete.
\end{theorem}

The positive direction (the existence of a $k$-Min-Max ordering
implies a polynomial time algorithm) is proved in \cite{yeo}.
We prove Theorem \ref{gut} using our characterization in
Theorem \ref{main-k}, by showing that MinHOM($H$)
is NP-complete if $H$ contains a sym-$k$-invertible pair
or an induced unbalanced oriented cycle of net length other
than $k$; this is done in the next section.

\section{The NP-completeness Claims}

Our basic NP-completeness tool is summarized in the next lemma.

\begin{lemma}\label{np-core}
Let $H$ be a digraph and $x, y$ two vertices of $H$; let $S$ be a
digraph and $s, t$ two vertices of $S$. Suppose we have costs
$c_j(i)$ of mapping vertices $i$ of $S$ to vertices $j$ of $H$
where $c_x(s)=c_x(t)=1, c_y(s)=c_y(t)=0$, and such that there
exists

\begin{itemize}
\item a homomorphism $f : S \rightarrow H$ mapping $s$ to $x$
and $t$ to $y$ of total cost $1$ (i.e., in which all other vertices of
$S$, different from $s, t$, map to vertices of $H$ with costs zero)

\item a homomorphism $g : S \rightarrow H$ mapping $s$ to $x$
and $t$ to $x$ of total cost $2$ (other vertices map with costs zero)

\item a homomorphism $h : S \rightarrow H$ mapping $s$ to $y$
and $t$ to $x$, of total cost $1$ (other vertices map with costs zero)

\item but no homomorphism $S \rightarrow H$ mapping $s$ to $y$
and $t$ to $y$ of cost at most $|V(S)|$.
\end{itemize}

Then MinHOM($H$) is NP-complete.
\end{lemma}

\pf Let $G$ be an arbitrary graph, an instance of the maximum
independent set problem. We construct a corresponding instance $D$
of MinHOM($H$) by replacing every edge of $G$ by a copy of $S$.
Note that $D$ contains all {\em old} vertices of $G$, as well as
the {\em new} vertices each lying in a separate copy of $S$. The
costs $c_i(j), i \in V(H), j \in V(D)$, are defined as follows.

\begin{itemize}
\item if $v$ is an old vertex of $G$, then $c_x(v)=1, c_y(v)=0$,
and $c_z(v)=|V(G)|$ for all other $z \in V(H)$,
\item if $v$ is a new vertex of $D$ lying in a copy of $S$, its costs
are determined by the corresponding costs $c_j(v)$ in $S$.
\end{itemize}

Note that since we have $c_x(s)=c_x(t)=1, c_y(s)=c_y(t)=0$, the
two parts of the definition don't conflict. We now claim that $G$
has an independent set of size $k$ if and only if there exists a
homomorphism of $D$ to $H$ of cost $|V(G)|-k$. Indeed, if $I$ is
an independent set in $G$, we define a homomorphism $\phi : D
\rightarrow H$ by setting $\phi(j)=y$ if $j \in I$, $\phi(j)=x$ if
$j \in V(G) \setminus I$, and extending this mapping to a
homomorphism of $D$ to $H$, using the mappings $f, g, h$. It is
clear that the cost of $\phi$ is exactly $|V(G)|-|I|$. Conversely, let
$f$ be any homomorphism of $D$ to $H$ of total cost less than
$|V(G)|$. Thus the old vertices of $G$ must map to either $x$ or
$y$. If two adjacent vertices are mapped to $y$ we incur a cost
of at least $|V(S)| \geq 2$. Thus we may assume that those vertices
that map to $y$ are independent. Since the old vertices of $G$ that
map to $x$ contribute a cost of one each, we conclude that if there
is a homomorphism of cost $|V(G)|-k$ then there is an independent
set of size $k$ in $G$. \qed

One example in which we can easily use this lemma deals with a
special case of sym-invertible pairs.

\begin{corollary}\label{one-avoid}
Suppose $u, v$ is a sym-invertible pair in $H$ with corresponding
walks $P, Q$, such that there are some faithful arcs from $P$
to $Q$ but there are no faithful arcs from $Q$ to $P$.

Then the problem MinHOM$(H)$ is NP-complete.
\end{corollary}

\pf We assume $P= u=a_1 \ldots  a_n=v, Q= v=b_1 \ldots b_n=u,$ and
let  $S = s_1 \ldots s_n$ be a {\em path} (all vertices are distinct)
congruent to $P$ (and $Q$). Define the cost of mapping vertices of
$S$ to $H$ as follows. If $c_{u}(s_1)=c_{u}(s_n)=1$, and
$c_{v}(s_1)=c_{v}(s_n)=0$, and $c_{a_i}(s_i)=c_{b_i}(s_i)=0$ for
$1 < i <n$. In any other case the cost is $n$.

Clearly there are obvious homomorphisms $\phi : S \rightarrow P$
and $\psi : S \rightarrow Q$. Define also $\zeta: S \rightarrow H$
to be the homomorphism defined by $\zeta(s_i)=a_i$ for $1 \leq i
\leq k$ and $\zeta(s_i)=b_i$ for $k+1 \leq i \leq n$. Let
$a_tb_{t+1}$ be a faithful arc from $P$ to $Q$. Suppose
there is homomorphism $g : V(S) \rightarrow V(P) \cup V(Q)$
such that $g(s_1)=g(s_n)=v$. Then the cost of $g$ is at least
$n$ unless $g(r_i)$ is $a_i$ or $b_i$. Since $g(s_1)=g(s_n)=v$,
there has to be a faithful arc from $Q$ to $P$ in $H$, which is
a contradiction. Now by apply Lemma \ref{np-core} for $P,Q$
and $S$ MinHOM($H$). \qed

We next consider the case where some sym-invertible pair has
faithful arcs both from $P$ to $Q$ and from $Q$ to $P$.

It was noted in \cite{mincostungraph} (using \cite{alekseevDM265})
that the following problem $\Pi_3$ is NP-complete. Given a
three-coloured graph $G$ and an integer $k$, decide if there
exists an independent set of $k$ vertices. It is easy to see that
this fact can be generalized to the following problem

$\Pi_{2m+1}$:

Given a graph $G$ with a homomorphism $f: G \rightarrow C_{2m+1}$,
decide if there exists an independent set of $k$ vertices.

\begin{lemma}
Each problem $\Pi_{2m+1}$ is NP-complete.
\end{lemma}

\pf Modify every instance $G$ of $\Pi_{2m-1}$ to an instance $G'$
of $\Pi_{2m+1}$ by replacing each edge of $G$ between classes
$f^{-1}(1)$ and $f^{-1}(2)$ by a path of length three. \qed

We apply this result as follows.

\begin{lemma}\label{two-avoid}
Suppose $u, v$ is a sym-invertible pair in $H$ with corresponding
walks $P, Q$, such that there are faithful arcs from $P$ to $Q$ as
well as faithful arcs from $Q$ to $P$.

Then MinHOM$(H)$ is NP-complete.
\end{lemma}

\pf The walks $P = x_0, x_1, \dots, x_n$ and $Q = y_0, y_1, \dots,
y_n$ can be organized into segments $P_1, \dots P_k, Q_1, \dots,
Q_k$, where for each $i$ all faithful arcs between $P$ and $Q$ go
from $P$ to $Q$ or from $Q$ to $P$. Assume $P_i=x_{r_{i-1}},
x_{r_{i-1}+1}, \dots, x_{r_i}$ and $Q_i=y_{r_{i-1}},
y_{r_{i-1}+1}, \dots, y_{r_i}$ with $r_0=0, r_k=n$, and assume,
without loss of generality, that there are faithful arcs from
$P_1$ to $Q_1$ but no faithful arcs from $Q_1$ to $P_1$, there are
faithful arcs from $Q_2$ to $P_2$ but no faithful arcs from $P_2$
to $Q_2$, etc. Note that if $k$ is odd, the faithful arcs of the
last segment go from $Q$ to $P$, and if $k$ is even, they go from
$P$ to $Q$. Let $R_i$ be a path congruent to $P_i$ (and $Q_i$),
and for simplicity assume that $R_i=r_{i-1}, \dots, r_i$.

{\bf Case 1. Assume $k$ is odd.}

We reduce $\Pi_k$ to MinHOM$(H)$ as follows. Consider an instance
of $\Pi_k$, namely, a graph $G$ with a homomorphism $f$ to $C_k$.
Suppose the vertices of $C_k$ are $1, 2, \dots, k$ (consecutively,
and viewed modulo $k$). Replace each edge $uv$ of $G$ with $u \in
f^{-1}(i)$ and $v \in f^{-1}(i+1)$ (modulo $k$) by a copy
$R_i(u,v)$ of $R_i$, identifying $r_{i-1}$ with $u$ and $r_i$ with
$v$, obtaining a digraph $D$. The costs of mapping an old vertex
(from $G$) $u$ in $f^{-1}(i)$ with $i$ odd will be
$c_{x_{r_i}}(u)=1, c_{y_{r_i}}(u)=0$, while the costs of mapping
an old vertex $u$ in $f^{-1}(i)$ with $i$ even will be
$c_{x_{r_i}}(u)=0, c_{y_{r_i}}(u)=1.$ For vertices inside the
substituted copies of $R$, we proceed as above, defining their
costs to be zero only for the corresponding vertices in $R(u,v)$.
All other costs are $|V(G)|$.

Suppose $i$ is odd. Each homomorphism of $R_i$ to $D$ taking
$r_{i-1}$ to $x_{r_{i-1}}$ and $r_i$ to $y_{r_i}$ has a very high
cost, but all other possibilities ($r_{i-1}$ to $x_{r_{i-1}}$ and
$r_i$ to $x_{r_i}$; $r_{i-1}$ to $y_{r_{i-1}}$ and $r_i$ to
$y_{r_i}$; and $r_{i-1}$ to $y_{r_{i-1}}$ and $r_i$ to $x_{r_i}$)
have cost $1$. A similar analysis applies to $i$ even. A special
consideration is needed for the last segment $R_k$, where we use
the fact that $x_{r_k}=x_n=y_0$ and $y_{r_k}=y_n=x_0$.

As in the proof of Corollary \ref{one-avoid}, these facts imply that
$G$ has an independent set of size $\ell$ if and only if $D$ has a
homomorphism to $H$ of cost $|V(G)|-\ell$.

{\bf Case 2. Assume $k$ is even.}

In this case instead of the sym-invertible pair $u, v$ with walks
$P, Q$ we consider the sym-invertible pair $y_{r_1}, x_{r_1}$ with
walks $P', Q'$ where $P'=y_{r_1}, \dots, y_{r_2}, \dots,
y_{r_{k-1}}, \dots y_{r_k}= y_n=x_0, \dots, x_{r_1}$, and
$Q'=x_{r_1}, \dots, x_{r_2}, \dots x_{r_{k-1}}, \dots
x_{r_k}=x_n=y_0, \dots, y_{r_1}$. Note that there are no faithful
arcs from $x_{r_{k-1}}, \dots x_{r_k}=x_n=y_0, \dots, y_{r_1}$ to
$y_{r_{k-1}}, \dots y_{r_k}=y_n=x_0, \dots, x_{r_1}$. Thus we
obtain an odd number of segments and we can proceed as above,
unless $k=2$ in which case we only have one segment and Corollary
\ref{one-avoid} applies. \qed

We can now handle the case when $H$ is balanced. Recall that
this means that the vertices of $H$ have levels $0, 1, \dots, h$ so that
each arc goes from some level $i$ to level $i+1$. It is easy to see
that in a balanced digraph a sym-invertible pair $u, v$ must have $u$
and $v$ on the same level. Thus all sym-$k$-invertible pairs have $k=1$,
i.e., we only have sym-invertible pairs. Therefore, the NP-completeness
part of Theorem \ref{gut} in this case reduces to the following.

\begin{theorem}\label{bal}
If a balanced digraph $H$ contains a sym-invertible pair, then
MinHOM$(H)$ is NP-complete.
\end{theorem}

\pf
By Corollary \ref{one-avoid} and Lemma \ref{two-avoid}, we may
assume that we have a sym-invertible pair $u, v$ and corresponding
walks $P, Q$ with no faithful arcs between $P$ and $Q$. Consider
the walk $W$ in $H^*$ from $(u,v)$ to $(v,u)$ corresponding to $P$
and $Q$. If some $(a,b)$ lies on $W$, then there is a walk in $H^*$
from $(a,b)$ to $(b,a)$ (because $H^*$ has an arc from $(x,y)$ to
$(x',y')$ if and only it has an arc from $(y,x)$ to $(y',x')$). Thus we
may assume that $u, v$ are on the lowest level of $P$ and $Q$.
Let $z$ be vertex on the highest level of $P$ and let $w$ be the
corresponding vertex on $Q$. Let $R$ be the walk obtained by
following $Q$ from $v$ to $w$ and then following $Q^{-1}$ back
from $w$ to $v$. Let the path $S$ be the common pre-image of
$P, Q$, and $R$, obtained by applying Lemma \ref{typer} twice,
since $P, Q, R$ consist of two constricted pieces. Let $f$ be the
corresponding homomorphism of $S$ to $P$, let $g$ be the
corresponding homomorphism of $S$ to $Q$, and let $h$ be
the corresponding homomorphism of $S$ to $R$. We define the
cost of mapping an internal vertex $j$ of $S$ to a vertex $i$ of $H$
to be zero if $ i \in \{f(j),g(j),h(j) \}$; the cost of mapping the first and
the last vertex of $S$ to $v$ is $1$ and to $u$ is zero. In all other
cases the cost is $|V(S)|$. Note that there is no homomorphism from
$S$ to $H$ which maps both beginning and end of $S$ to $u$ of
total cost smaller than $|V(S)|$, as otherwise there would be a faithful
arc from $P$ to $Q$. Now by applying Lemma \ref{np-core} to $S$
and $f, g, h$ we conclude that MinHOM($H$) is NP-complete. \qed

\begin{corollary}\label{none-avoid-balanced}
Theorem \ref{gut} holds for balanced digraphs $H$.

Specifically, for a balanced digraph $H$ the problem MinHOM$(H)$
is polynomial time solvable if $H$ has a Min-Max ordering, and is
NP-complete otherwise.
\end{corollary}

We observe that the same proof applies even in unbalanced digraphs
$H$ as long as $P$ (and hence $Q$) has net length zero. Specifically,
if {\em any} digraph $H$ has an invertible pair $u, v$ with corresponding
walks $P, Q$ which have net length zero, then MinHOM$(H)$ is NP-complete.

Thus we may now focus on unbalanced digraphs $H$.

\begin{theorem}\label{two-oriented-cycle}
Suppose $H$ is weakly connected and contains two induced oriented cycles
$C_1,C_2$, with net lengths $k, n > 0, k \ne n$.

Then MinHOM($H$) is NP-complete.
\end{theorem}

We will use the following analogue of Lemma \ref{typer}
for infinite walks which are  constricted in the infinite sense,
i.e., are constricted from below and have infinite height.

\begin{corollary}\label{typeinf}
Let $P_1$ and $P_2$ be two walks of infinite height, constricted
from below. Assume that $P_i$ starts in $p_i$, $i=1, 2$, and let
$q_i$ be a vertex on $P_i,$ such that the infinite portion of $P_i$
starting from $q_i$ is also constricted from below, and the portions
of $P_i$ from $p_i$ to $q_i$ have the same net length, for $i=1, 2$.

Then there is an oriented path $P$ that admits homomorphisms $f_i$
to $P_i$ taking the starting vertex of $P$ to $p_i$ and the ending vertex
of $P$ to $q_i$, for $i=1, 2$.
\end{corollary}

{\bf Proof of the Corollary:}
Let $P'_i$ be the portion of $P_i$ from $p_i$ to $q_i$, and suppose,
without loss of generality, that the height $h$ of $P'_1$ is greater than
or equal to the height of $P'_2$. Let $r_i$ be the first vertex after $q_i$.
(or equal to $q_i$) on $P_i$, such that the net length from $p_i$ to $r_i$
is $h$. Let $R_i$ be the subwalk of $P_i$ from $p_i$ to $r_i$. Now
Lemma \ref{typer} implies that there is a path $R$ with homomorphisms
$f_i$ to $R_i$ taking the beginning of $R$ to $p_i$ and the end of $R$
to $r_i$. Suppose $x$ is the last vertex on $P'_1$ with $f_1(x)=q_1$: if
$f_2(x)=q_2$, we are done, so suppose $f_2(x)=y \neq q_2$. Now
consider the subwalk $Y$ of $P'_2$ joining $y$ and $q_2$: it has net
length zero and is constricted from below, because the portion of $R$
between $x$ and the end of $R$ has net length zero and is constricted
from below. Let $h'$ be the height of $Y$, and let $X$ be the walk on
$P'_1$ from $q_1$ to the first vertex making a net length $h'$ and then
back to $q_1$. Since $X$ and $Y$ have the same height and have net
length zero, we can split them into two constricted pieces, and so Lemma
\ref{typer} implies that there is a path $R'$ which is a common pre-image
of $X$ and $Y$. Concatenating $R$ with $R'$ yields a path $P$ and
we can extend the homomorphisms $f_i$ to $P$ so that also the ending
vertex of $P$ is taken to $q_i$, for $i=1, 2$.
\qed

{\bf Proof of the Theorem:}
Suppose $k > n$, so $k$ does not divide $n$. We may assume that
$H$ is minimal, in the sense that no weakly connected subgraph $H'$
of $H$ with fewer vertices contains two induced cycles with different
non-zero net lengths. Indeed, if $H'$ were such a subgraph, then
MinHOM($H'$) would be polynomially reduced to MinHOM($H$) by
the cost of mapping to vertices of $H$ not in $H'$ very high.

Each cycle $C_i, i=1,2$, contains a vertex $u_i$ such that the
walk starting in $u_i$ and following $C_i$ (in the positive direction)
is constricted from below. Let $U$ be a walk in $H$ from $u_1$ to $u_2$,
and let $u$ be a vertex on $U$ of minimum height. By minimality, we
may assume $V(H)=V(C_1) \cup V(C_2) \cup U$. Let $P_i, i=1, 2$,
be the walk from $u$ to $u_i$ following $U$ (or $U^{-1}$), then once
around $C_i$ (in the positive direction), and then back from $u$
following $U^{-1}$ (or $U$). It follows that each $P_i$ is constricted
from below. The net length of $P_1$ is $k$ and the net length of
$P_2$ is $n$. Let $Q_i$, $i=1, 2$, be the infinite walk starting at $u$
obtained by repeatedly concatenating $P_i$, and let $Q'_i$ be the
two-way infinite walk obtained by expanding $Q_i$ in the opposite
direction by repeatedly concatenating $P^{-1}_i$.

Let $d$ be greatest common divisor of $n$ and $k$, and let $a=k/d - 2$.
Thus $(a+2)n$ is the smallest positive common multiple of $n$ and $k$.
We now define the following three walks $W_1, W_2, W_3$ in $H$ of
net length $(a+1)n$.

\begin{enumerate}
\item
The walk $W_1$ starts at $u$ and follows $Q_1$ going around $P_1$
until the last vertex $v$ such that the net length of the resulting walk is
$(a+1)n$
\item
$W_2$ also starts at $u$ and follows $Q_2$ going around $P_2$ fully
$(a+1)$ times, ending at $u$
\item
$W_3$ starts at $v$ and follows $P_1$ until the first occurrence of $u$,
and then continues $a$ times around $P_2$, ending again at $u$.
\end{enumerate}

Now we define, in analogy with $Q_1, Q_2$, also the infinite walk
$Q_3$, obtained from $W_3$ by continuing to go around $P_2$.
Because we chose $v$ to be the last on $Q_1$ with the right net length,
the walk $W_3$ is constricted from below; of course $W_1, W_2$ are
also constricted from below. Hence $Q_1, Q_2, Q_3$ are also constricted
from below; they have infinite heights because $C_1, C_2$ have positive
net length. Thus we can apply Corollary \ref{typeinf} to $Q_1, Q_2, Q_3$,
obtaining a common pre-image which is a path $S$, say $s=s_0, s_1,$
$\dots, s_q=t$, with homomorphisms $f, g, h$ of $S$ to $Q_1, Q_2, Q_3$
respectively, such that

\begin{enumerate}
\item $f(s)=u, f(t)=v$
\item $g(s)=g(t)=u$
\item $h(s)=v, h(t)=u$
\end{enumerate}

Note that the walk $W'_1$ equal to $u=f(s_0), f(s_1),$ $\dots, f(s_q)=v$,
the walk $W'_2$ equal to $v=g(s_0), g(s_1),$ $\dots, g(s_q)=u$, and the
walk $W'_3$ equal to $v=h(s_0), h(s_1),$ $\dots, h(s_q)=u$ are congruent.

Assume first that $W'_1, W'_3$ do not avoid each other, i.e., for some $i$
we have both the faithful arcs (forward or backward) $f(s_i)h(s_{i+1}),$
$h(s_i)f(s_{i+1})$. Note that $W'_1 \cup W'_2 \cup W'_3$ contains all the
vertices of $H$, so the minimality of $H$ easily implies that all four vertices
$f(s_i), h(s_i), f(s_{i+1}), h(s_{i+1})$ must belong to $C_1 \cup C_2$. Since
the cycles are induced. we must have two vertices in each cycle. Up to
symmetry, we may assume we have forward arcs $ab \in C_1$ and
$cd \in C_2$, as well as forward arcs $ad, cb$ in $H$. Then, say,
$a=f(s_i)$, $b=f(s_{i+1})$, $c=h(s_i)$, $d=h(s_{i+1})$.

We first claim that $C_1, C_2$ do not have common vertices, or
arcs joining them other than $ad, cb$. Otherwise, let $x$ on $C_1$
be the first vertex following $b$ in the direction opposite to $a$,
equal to or adjacent with some $y$ on $C_2$, and assume that
$y$ is the first vertex of $C_2$ following $d$, in the direction
opposite to $c$, adjacent to $x$. Consider the cycle
$D_1$ with arcs $ab, ad, xy$, the portion of $C_1$ between $b$ and
$x$ not containing $a$, and the portion of $C_2$ between $d$ and
$y$ not containing $c$, and the cycle $D_2$ with arcs $cb, cd,
xy$, and the same portions of $C_1, C_2$. The cycles $D_1, D_2$
have the same net length $m$. If $m$ is not zero and not $k$, we
could delete $c$ and obtain a smaller weakly connected $H'$ with
two different non-zero net lengths. If $m$ is not zero and not $n$
we could likewise delete $a$. Thus $m=0$. If $x$ has no
neighbours on $C_2$ other than $y$, then consider instead of
$D_2$ the cycle $D_2'$ obtained from $D_2$ by replacing the
portion of $C_2$ between $c$ and $y$ containing $d$ by the
portion of $C_2$ between $c$ and $y$ not containing $d$. Since
$m=0$, the net length of $D'_2$ is $n$, so we can delete $d$ and
obtain a smaller weakly connected $H'$ with two different non-zero
net lengths. Otherwise, let $y_1, y_2, \dots, y_p$ be all the neighbours
of $x$ on $C_2$ after $y=y_0$, numbered consecutively in the direction
from $y$ to $c$, away from $d$. Consider the oriented cycles $Y_i$
containing $x, y_i, y_{i+1}$ and the segment of $C_2$ between
$y_i$ and $y_{i+1}$ not containing $d$. Each $Y_i$ is an induced
cycle in $H$, and the sum of their net lengths in $n$. Hence at least
one $Y_i$ has a non-zero net length and we similarly obtain a
contradiction with the minimality of $H$.

Thus $H$ consists of $C_1, C_2$, and the two extra arcs (forward
or backward) $ad, cb$; in particular $u \in C_1 \cup C_2$, and the
path $U$ uses $ad$ or $bc$. Without loss of generality, we may
assume that it uses $bc$, since we can replace $ad$ by $ab, bc,
cd$. Suppose first that $u \in C_1$, whence we also have $v \in
C_1$. Consider the initial portion of $W'_1$ from $v$ to
$b=f(s_{i+1})$: it has net length equal to a multiple of $k$
(corresponding to going full rounds around the cycle $C_1$) plus
the net length of the portion $X_1$ of $C_1$ (in the positive
direction) from $u$ to $b$. Consider next the initial portion of
$W'_3$ from $v$ to $c$ followed by the arc joining $c$ and $b$: it
has net length equal to $n$ (corresponding to going from $v$ to
$u$, which must precede $c \in C_2$) plus a multiple of $n$
(corresponding to going full rounds around the closed walk $P_2$
from $u$ to $u$) plus the net length of the portion $X_2$ of $P_2$
(in the positive direction) from $u$ to $c$ concatenated with the
arc joining $c$ and $b$. However, from $u$ to $c$ we must use the
arc joining $b$ and $c$. Thus $X_2$ uses the arc joining $b$ and
$c$ first in one direction and then in the opposite direction,
whence the net lengths of $X_1, X_2$ are the same. This means that
a multiple of $n$, smaller than $(a+2)n$ is also a multiple of
$k$, which is impossible.

It remains to consider the case when $W'_1, W'_3$ do avoid
each other. We now assume that of all homomorphisms $f, g, h$
of $S$ to $Q_1, Q_2, Q_3$ satisfying properties (1, 2, 3) and
such that the resulting walks $W'_1, W'_3$ avoid each other,
we have chosen ones that maximize the number of vertices
with $f(s_i)=g(s_i)$ or $g(s_i)=h(s_i)$.

If $W'_1, W'_3$ have at least some faithful arcs, then Corollary
\ref{one-avoid} and Lemma \ref{two-avoid} imply MinHOM$(H)$
is NP-complete. Thus we may assume that there are no faithful
arcs between $W'_1$ and $W'_3$.

We now define the costs of mapping vertices $x$ of $S$ to vertices
$j$ of $H$ as follows: $c_j(x)=|S|$ except for $c_u(s)=c_u(t)=1$,
$c_v(s)=c_v(t)=0$ and $c_j(s_i)=0$ when $j \in \{f(s_i),g(s_i),h(s_i)\}$,
$j \neq u$.

By properties (1, 2, 3), we see that to apply the Lemma \ref{np-core}
it remains to show that there is no homomorphism of $S$ to $H$ of
cost $|S|-1$ or less, taking both $s$ and $t$ to $v$. Suppose, for a
contradiction, that there is such a homomorphism $\phi$. Then we
must have $\phi(s_0)=h(s_0)$, $\phi(s_q)=f(s_q)$, and each
$\phi(s_i) \in \{f(s_i), g(s_i), h(s_i)\}$. Since there are no faithful
arcs between $W'_1$ and $W'_3$, we can't have $h(s_i)$ and
$f(s_{i+1})$ adjacent. Thus, because of the costs, we must have
some $h(s_i)$ and $g(s_{i+1})$ as well as $g(s_j)$ and $f(s_{j+1})$
are adjacent, with $i<j$. We now claim that this contradicts the
maximality of $f, g, h$. Indeed, we could redefine $f$ to equal
$g$ up to $s_j$ (and then continuing as before, taking advantage
of the arc joining $g(s_j)$ and $f(s_{j+1})$), obtaining a new
$W'_1$ with at least one more vertex (namely $s_{i+1}$) having
equality of $f$ and $g$. (We need to observe that the new $W'_1$
still avoids $W'_3$, which also follows by maximality of $f, g, h$:
there cannot be an arc between $g(s_p) \neq h(s_p)$ and $h(s_{p+1})$.)

From the theorem we also derive the following corollary
that will complete the proof of Theorem \ref{gut}.

\begin{theorem}\label{k-sym}
Suppose $H$ is a digraph containing an induced oriented cycle of
net length $k >0$. If there is homomorphism $\ell : H \rightarrow
\vec{C}_k$ with a sym-$k$-invertible pair, then MinHOM$(H)$ is
NP-complete.
\end{theorem}

\pf Recall that $P$ is a walk from $u$ to $v$ and $Q$ a congruent
walk with $P$, from $v$ to $u$. Recall also that there is a
homomorphism $\ell : H \rightarrow \vec{C}_k$, and
$\ell(u)=\ell(v)$. It follows that the net length of $P$ (and of
$Q$) is divisible by $k$. If there are faithful arcs from $P$ to
$Q$ or from $Q$ to $P$ then by Corollary \ref{one-avoid} or
\ref{two-avoid}, MinHOM$(H)$ is NP-complete. So we may assume
that there are such faithful arcs. We may assume that the net length
of $P$ is greater than zero as otherwise remark following Lemma
\ref{none-avoid-balanced} implies that MinHOM($H$) is NP-complete.
We now proceed to find congruent walks from $u$ to $v$ and from
$v$ to $u$ which avoid each other, and another congruent walk from
$u$ to $u$, so that we can apply Lemma \ref{np-core} in a fashion
similar to what was done in the proof of Theorem \ref{two-oriented-cycle}.

We may assume that $P$ is constricted from below, as otherwise we
replace $u,v$ by vertices $u' \in P$, $v' \in Q$, where $u'$ is a
vertex of $P$ with the minimum height, and $v'$ is the corresponding
vertex of $v'$ in $Q$. We have observed that $u', v'$ is also a
sym-$k$-invertible pair, thus there are walks $P'$ from $u'$ to
$v'$ and $Q'$ from $v'$ to $Q'$ that avoid each other. It is easy
to see that the minimality of $u'$ implies that this new $P'$ is
constricted from below. Let $C$ be a walk in $H$ from $u$
to an oriented cycle of net length $k$, followed by going around
the oriented cycle once in the positive direction and then
returning back on the same walk to $u$. Note that the net
length of this walk is $k$. We may again assume that $C$
is constricted from below, as otherwise instead of $P, Q$ we
could use $P_1, Q_1$, where $P_1$ is obtained by concatenating
$P$ with $(QP)^a$  and $Q_1$ is obtained by concatenating $Q$
with $(PQ)^a$ for some positive $a$, such that the walk from $u$
(at the beginning of $P_1$) to the $(a-1)$-th appearance of $u$ in
$P_1$, followed by $C$ is a walk constricted from below.

Let the net length of $P$ be $\ell k$, with $\ell > 0$. Let $W$ be the
infinite walk obtained by repeatedly concatenating $C$; note that
$W$ is constricted from below. Let $P'$ be the infinite walk obtained
by concatenating $P$ with infinitely many repetitions of $QP$. Let
$Q'$ be the infinite walk congruent to $P'$ obtained by similarly
concatenating $Q$ with repetitions of $PQ$. Let $C'$ be the walk
in $W$, from $u$ to a vertex $u'$ that is the $\ell$-th occurrence
of $u$ in $W$. Now we apply Corollary \ref{typeinf} to obtain a path
$S=s_0, s_1, \dots, s_t$ which is the common pre-image of $P, C', Q$.
In this application, we use $P', W, Q'$ as the infinite walks, and the
ends of $P,C', Q$ as the vertices $q_i$. (Note that $P, C', Q$ all have
net length $\ell k$. Corollary \ref{typeinf} also yields homomorphisms
$f, g, h$ of $S$ to $P', W, Q'$ taking $s_0$ to the beginnings of $P', W, Q'$
(also the beginnings of $P, C', Q$), and taking $s_t$ to the ends of
$P, C', Q$. Let $P''$ be the walk $f(s_0), f(s_1), \dots, f(s_t)$, let
$Q''$ be the walk $h(s_0), h(s_1), \dots, h(s_t)$, and let
$C''$ be the walk $g(s_0), g(s_1), \dots, g(s_t)$. Observe that
$P'', Q''$ avoid each other and between the walks $P'', Q''$ there
are no faithful arc, because that was the case for $P, Q$.

Note that $f(s_0)=u$ and $f(s_t)=v$, $g(s_0)=g(s_t)=u$
and $h(s_0)=v,h(s_t)=u$. We define the costs as follows, the
$c_u(s_0)=c_u(s_t)=1$, and $c_v(s_0)=c_v(s_t)=0$, and $c_i(x)=0$ when $i
\in \{f(x),g(x),h(y)\}$, $x \ne u$. For any other case the cost is
$|V(S)|$.

We now conclude the proof as in Theorem \ref{two-oriented-cycle},
assuming that the homomorphisms $f, g, h$ of $S$ to
$V(P'') \cup V(C'') \cup V(Q'')$ satisfy properties 1, 2, 3, and
maximize the number of vertices with $f(s_i)=g(s_i)$ or $g(s_i)=h(s_i)$.
\qed

We are finally ready to conclude the {\bf Proof of Theorem \ref{gut}},
i.e., to prove the conjecture from \cite{yeo}.

Recall that the polynomial case of the Theorem has been established
in \cite{yeo}. For the NP-completeness claim, the case when $H$ is balanced
in handled by Corollary \ref{none-avoid-balanced}. Thus we may assume
that $H$ has an induced oriented cycle of some positive net length $k$.
It is a well-known fact (e.g. Corollary 1.17 in \cite{hombook}) that $H$ has
a homomorphism to $\vec{C}_k$ if and only if it does not contain a closed
walk of net length not divisible by $k$. Suppose first that $H$ does not
admit a homomorphism to $\vec{C}_k$. Then the above fact implies that
$H$ contains an induced oriented cycle of net length not divisible by $k$.
Hence the problem MinHOM$(H)$ is NP-complete by Theorem
\ref{two-oriented-cycle}. If, on the other hand, $H$ does admit a
homomorphism to $\vec{C}_k$, with a sym-$k$-invertible pair,
then MinHOM$(H)$ is NP-complete by Theorem \ref{k-sym}. This
completes the proof.


\begin{thebibliography}{99}

\bibitem{archiv} R.S. Takhanov, Dichotomy theorem for general minimum cost
homomorphism problem, arXiv:0708.326v5.

\bibitem{alekseevDM265} V.E. Alekseev and V.V. Lozin, Independent sets of maximum
weight in $(p,q)$-colorable graphs, {\em Discrete Mathematics} 265 (2003) 351--356.

\bibitem{bang95} J. Bang-Jensen, P. Hell, and G. MacGillivray, Hereditarily hard
H-colouring problems, {\em Discrete Math.} 138 (1995) 75--?92.

\bibitem{barto1} L. Barto, M. Kozik, and T. Niven, The CSP dichotomy holds for
digraphs with no sources and sinks (a positive answer to a conjecture of
Bang-Jensen and Hell), {\em SIAM J. Computing}, in print.

\bibitem{barto2} L. Barto, M. Kozik, M. Mar\' oti, and T. Niven, CSP dichotomy for
special triads, {\em Proc. AMS}, in print.

\bibitem{bula} A. A. Bulatov,
Tractable conservative constraint satisfaction problems,
18th IEEE SOLCS (2003) pp. 321--330.

\bibitem{kroch} A. A. Bulatov, P. G. Jeavons, and A. A. Krokhin,
Constraint satisfaction problems and finite algebras,
27th ICALP (2000) pp. 272--282.

\bibitem{cater} C Carvalho, V. Dalmau, A. Krokhin,
Caterpillar duality for constraint satisfaction problems,
23rd Annual IEEE Symposium LICS, 307-316.

\bibitem{cohen} D. Cohen, M. C. Cooper, P. Jeavons,
Constraints, consistency and closure, Artificial Intelligence 101 (1998) 251Ð265.

\bibitem{cohenJAIR22}  D. Cohen, M. Cooper, P. Jeavons, and A. Krokhin,
A maximal tractable class of soft constraints. {\em J. Artif. Intell. Res.} 22 (2004) 1--22.

\bibitem{jonsson} T. F\" arnqvist and P. Jonsson,
Bounded treewidth and CSP-related problems, ISAAC2007.

\bibitem{pavol} T. Feder, P. Hell, and J. Huang,
Bi-arc graphs and the complexity of list homomorphisms, {\em J.
Graph Theory} 42 (2003) 61 - 80.

\bibitem{lists} T. Feder, P. Hell, J. Huang, and A. Rafiey,
Adjusted interval digraphs, Electronic Notes in Discrete Mathematics
32 (2009) 83--91.

\bibitem{fv} T. Feder, M. Vardi,
The computational structure of monotone monadic SNP and constraint
satisfaction: a study through datalog and group theory, {\em SIAM Journal
on Computing} 28 (1998) 57--104.

\bibitem{arv} A. Gupta, P. Hell, M. Karimi, and A. Rafiey,
Minimum cost homomorphisms to reflexive digraphs,
LATIN 2008.

\bibitem{mincostungraph} G. Gutin, P. Hell, A. Rafiey and A. Yeo,
A dichotomy for minimum cost graph homomorphisms,
{\em European J. Combin.} 29 (2008) 900--911.

\bibitem{gutinDAM} G. Gutin, A. Rafiey and A. Yeo, Minimum cost
and list homomorphisms to semicomplete digraphs, {\em Discrete
Appl. Math.} 154 (2006), 890--897.

\bibitem{yeo} G. Gutin, A. Rafiey and A. Yeo, Minimum Cost
Homomorphisms to Semicomplete Bipartite Digraphs, {\em SIAM
Journal of Discrete Math.} 22 (2008) 1624--1639.

\bibitem{arash} G.Gutin, A. Rafiey, A.Yeo, Minimum Cost Homomorphism
to Oriented Cycles. Submitted.

\bibitem{gutinDAMlora} G. Gutin, A. Rafiey, A. Yeo and M. Tso, Level of
repair analysis and minimum cost homomorphisms of graphs. {\em
Discrete Appl. Math.} 154 (2006) 881--889.

\bibitem{don} R. H\:aggkvist, P. Hell, D. J. Miller, and V. Neumann Lara,
On multiplicative graphs and the product conjecture, Combinatorica
8 (1988) 71Ð81.

\bibitem{halld2001} M. M. Halldorsson, G. Kortsarz, and H. Shachnai,
Minimizing average completion of dedicated tasks and interval
graphs. Approximation, Randomization, and Combinatorial
Optimization (Berkeley, Calif, 2001), Lecture Notes in Computer
Science, vol. 2129, Springer, Berlin, 2001, pp. 114--126.

\bibitem{hellJCT48} P. Hell and J. Ne\v{s}et\v{r}il, On the complexity
of $H$-colouring, {\em J. Combin. Theory B} 48 (1990) 92--110.

\bibitem{hombook} P. Hell and J. Ne\v{s}et\v{r}il, {\em Graphs and
Homomorphisms.} Oxford University Press, Oxford, 2004.

\bibitem{jansenJA34} K. Jansen, Approximation results for the
optimum cost chromatic partition problem. {\em J. Algorithms} 34
(2000) 54--89.

\bibitem{jeav} P. G. Jeavons,
On the algebraic structure of combinatorial problems,
Theoret. Comput. Sci. 200 (1998) 185--204.

\bibitem{jiangGT32} T. Jiang and D. B. West, Coloring of trees with
minimum sum of colors. {\em J. Graph Theory} 32 (1999) 354--358.

\bibitem{kroon1997} L.G. Kroon, A. Sen, H. Deng, and A. Roy, The optimal
cost chromatic partition problem for trees and interval graphs,
Graph-Theoretic Concepts in Computer Science (Cadenabbia, 1996),
Lecture Notes in Computer Science, vol. 1197, Springer, Berlin,
1997, pp. 279--292.

\bibitem{spin} J. Spinrad, {\bf Efficient Graph Representations},
Fields Institute Monographs, AMS 2003.

\bibitem{supowitCAD6} K. Supowit, Finding a maximum planar subset of a set
of nets in a channel. {\em IEEE Trans. Computer-Aided Design} 6
(1987) 93--94.

\bibitem{zhu} X. Zhu, A simple proof of the multipilicatively of
directed cycles of prime power length. {\em Discrete Appl. Math.}
36 (1992) 333-345.

\end{thebibliography}
\end{document}